\documentclass[sn-nature,pdflatex,12pt]{sn-jnl}

\usepackage{graphicx}%
\usepackage{multirow}%
\usepackage{amsmath,amssymb,amsfonts}%
\usepackage{amsthm}%
\usepackage{mathrsfs}%
\usepackage[title]{appendix}%
\usepackage[dvipsnames]{xcolor}
\usepackage[table]{xcolor} 
\usepackage{textcomp}%
\usepackage{longtable}
\usepackage{hyperref}
\usepackage{array, colortbl}
\usepackage{tcolorbox}
\usepackage{manyfoot}%
\usepackage{booktabs}%
\usepackage{algorithm}%
\usepackage{algorithmicx}%
\usepackage{algpseudocode}%
\usepackage{listings}%
\usepackage[nolist]{acronym}
\usepackage{comment}
\usepackage{tikz}
\usepackage{eurosym}
\usepackage{array} 
\usepackage{enumitem}
\usepackage{lscape}
\usepackage{pgfplots}
\usepackage{supertabular}
\usepackage{multirow}
\usepackage{lscape}
\usepackage{tabularx}
\usepackage{xr}
\usepackage{tikz}
\usepackage{lipsum}
\usepackage{soul}
\usepackage{makecell}
\usepackage{CJKutf8}
\usepackage{xurl}
\usepackage{lineno}

\newcounter{mybox}
\renewcommand{\themybox}{Box \arabic{mybox}}
\renewcommand{\figurename}{Fig.}

\graphicspath{{graphics/}}
\pgfplotsset{compat=1.17}
\definecolor{nature_red}{RGB}{166,25,46}
\definecolor{nature_gray}{RGB}{245,245,245}

\theoremstyle{thmstyletwo}%

\theoremstyle{thmstylethree}%

\renewcommand{\figurename}{FIG.}

\begin{document}

\title[\centering Towards socio-techno-economic power systems \\ with demand-side flexibility]{\centering Towards socio-techno-economic power systems \\ with demand-side flexibility}


\author[1]{\fnm{Hanmin} \sur{Cai}}\email{hanmin.cai@empa.ch}
\equalcont{These authors contributed equally to this work.}

\author[1]{\fnm{Federica} \sur{Bellizio}}\email{federica.bellizio@empa.ch}
\equalcont{These authors contributed equally to this work.}

\author*[2]{\fnm{Yi} \sur{Guo}}\email{yi.guo@ieee.org}
\equalcont{These authors contributed equally to this work.}

\author[1]{\fnm{Gabriele} \sur{Humbert}}\email{gabriele.humbert@empa.ch}

\author[1]{\fnm{Mina} \sur{Montazeri}}\email{mina.montazeri@empa.ch}

\author[1,3]{\fnm{Julie} \sur{Rousseau}}\email{julie.rousseau@eeh.ee.ethz.ch}

\author[1]{\fnm{Matthias} \sur{Brandes}}\email{matthias.brandes@empa.ch}

\author[1]{\fnm{Arnab} \sur{Chatterjee}}\email{arnab.chatterjee@empa.ch}

\author[1,4]{\fnm{Andrea} \sur{Gattiglio}}\email{andrea.gattiglio@epfl.ch}

\author[1,5]{\fnm{Leandro} \sur{von Krannichfeldt}}\email{leandro.vonkrannichfeldt@empa.ch}

\author[1]{\fnm{Emmanouil} \sur{Thrampoulidis}}\email{mthrambo@gmail.com}

\author[1,6]{\fnm{Varsha} \sur{N. Behrunani}}\email{varsha.behrunani@empa.ch}

\author[7]{\fnm{Goran} \sur{Strbac}}\email{g.strbac@imperial.ac.uk}

\author[1]{\fnm{Philipp} \sur{Heer}}\email{philipp.heer@empa.ch}

\affil[1]{\orgdiv{Urban Energy Systems Laboratory}, \orgname{Swiss Federal Laboratories for Materials Science and Technology (Empa)}, \orgaddress{\street{Ueberlandstrasse 129}, \city{Dübendorf}, \postcode{8600}, \state{Zürich}, \country{Switzerland}}}

\affil[2]{\orgdiv{School of Automation}, \orgname{Beijing Institute of Technology}, \orgaddress{\city{Beijing}, \country{China}}}

\affil[3]{\orgdiv{Power Systems Laboratory}, \orgname{ETH Zürich}, \orgaddress{\street{Physikstrasse 3}, \city{Zürich}, \postcode{8092}, \state{Zürich}, \country{Switzerland}}}

\affil[4]{\orgdiv{Automatic Control Laboratory}, \orgname{EPFL}, \orgaddress{\street{Route Cantonale}, \city{Lausanne}, \postcode{1015}, \state{Vaud}, \country{Switzerland}}}

\affil[5]{\orgdiv{Intelligent Maintenance and Operations Systems Laboratory}, \orgname{EPFL}, \orgaddress{\street{Route Cantonale}, \city{Lausanne}, \postcode{1015}, \state{Vaud}, \country{Switzerland}}}

\affil[6]{\orgdiv{Automatic Control Laboratory}, \orgname{ETH Zürich}, \orgaddress{\street{Physikstrasse 3}, \city{Zürich}, \postcode{8092}, \state{Zürich}, \country{Switzerland}}}

\affil[7]{\orgdiv{Control and Power Research Group}, \orgname{Imperial College London}, \orgaddress{\city{London}, \postcode{SW7 2AZ}, \country{United Kingdom}}}


\abstract{\normalsize 
Harnessing the demand-side flexibility in building and mobility sectors can help to better integrate renewable energy into power systems and reduce global CO\textsubscript{2} emissions. Enabling this sector coupling can be achieved with advances in energy management, business models, control technologies, and power grids. The study of demand-side flexibility extends beyond engineering, spanning social science, economics, and power and control systems, which presents both challenges and opportunities to researchers and engineers in these fields. This Review outlines recent trends and studies in social, economic, and technological advancements in power systems that leverage demand-side flexibility. We first provide a concept of a socio-techno-economic system with an abstraction of end-users, building and mobility sectors, control systems, electricity markets, and power grids. We discuss the interconnections between these elements, highlighting the importance of bidirectional flows of information and coordinated decision-making. We then emphasize that fully realizing demand-side flexibility necessitates deep integration across stakeholders and systems, moving beyond siloed approaches. Finally, we discuss the future directions in renewable-based power systems and control engineering to address key challenges from both research and practitioners' perspectives. A holistic approach for identifying, measuring, and utilizing demand-side flexibility is key to successfully maximizing its multi-stakeholder benefits but requires further transdisciplinary collaboration and commercially viable solutions for broader implementation. 
}


\maketitle


\begin{acronym}[ML] 
\acro{AC}{Air Conditioning}
\acro{AI}{Artificial Intelligence}
\acro{ANN}{Artificial Neural Networks}
\acro{XAI}{Explainable Artificial Intelligence}
\acro{ML}[ML]{Machine Learning}
\acro{MPC}[MPC]{Model Predictive Control}
\acro{RBC}[RBC]{Rule Based Control}
\acro{DER}{Distributed Energy Resource}
\acro{HP}{Heat Pump}
\acro{IEA}{International Energy Agency}
\acro{ICT}{Information and Communication Technology}
\acro{LFM}{Local Flexibility Market}
\acro{LIME}{Local Interpretable Model-agnostic Explanations}
\acro{SHAP}{SHapley Additive exPlanations}
\acro{EV}{Electric Vehicle}
\acro{DSO}{Distribution System Operator}
\acro{TSO}{Transmission System Operator}
\acro{ADN}{Active Distribution Network}
\acro{DSM}{Demand Side Management}
\acro{DR}{Demand Response}
\acro{MILP}{Mixed-Integer Linear Programming}
\acro{MDP}{Markov Decision Process}
\acro{DeePC}{Data-enabled Predictive Control}
\acro{P2P}{Peer-to-Peer}
\acro{RL}{Reinforcement Learning}
\acro{FLC}{Fuzzy Logic Control}
\acro{CHP}{Combined Heat and Power}
\acro{PID}{Proportional-Integral-Derivative}
\acro{DRL}{Deep Reinforcement Learning}
\acro{KPI}{Key Performance Indicator}
\acro{HVAC}{Heating, Ventilation and Air Conditioning}
\acro{OCP}{Optimal Control Problem}
\acro{PID}{Proportional–integral–derivative}
\acro{TES}{Thermal Energy Storage}
\acro{DHN}{District Heating Network}
\acro{PV}{Photovoltaics}
\acro{SH}{Space Heating}
\acro{DHW}{Domestic Hot Water}
\acro{EMS}{Energy Management System}
\acro{TDNN}{Time-Delayed Neural Network}
\acro{RT}{Regression Tree}
\acro{V2G}{Vehicle-to-Grid}
\acro{PCM}{Phase Change Materials}
\acro{RES}{Renewable Energy Resources}
\end{acronym}

\newpage
\section*{Introduction}\label{sec:introduction}
Global electricity demand is expected to double by 2050, reaching between 52,000 and 71,000~TWh~\cite{mckinsey2023energy}. This substantial increase is driven by population growth, economic development, and the widespread electrification of mobility, buildings, and industry sectors. To meet this demand while adhering to global decarbonization targets, renewable energy resources are expected to contribute 85\% of global power generation by 2050, with solar and wind playing dominant roles \cite{irena2018get}. However, the large-scale integration of intermittent renewable energy sources poses operational challenges, particularly in maintaining reliability amid rising loads and supply fluctuations.\\

Demand-side flexibility has emerged as a promising solution for enhancing power system operation and supporting the integration of renewables \cite{Johanna2024_task_force,callaway2010achieving}. Demand-side flexibility is the capacity of consumers to adjust their load profiles without compromising their comfort and productivity. Two key demand-side sectors, including building and mobility, accounted for 29\% of global final energy consumption in 2019 \cite{IEA2023_Extended_world_energy_balances} and contributed 26\% of greenhouse gas emissions \cite{minx2021comprehensive}. By 2050, 70\% of space and water heating is expected to be electrified, and fuel/technology mandates will drive the full electrification of passenger vehicles by 2040 \cite{van2025demand}. This transition amplifies the potential of demand-side flexibility to mitigate grid stress, reduce emissions, and enhance system efficiency \cite{langevin2023demand}.  \\

A well-designed \ac{DSM} system can harness demand-side flexibility to optimize the power consumption of end-users, offering a flexible counterbalance to the uncertainty of renewable generations~\cite{wang2018transactive}. For example, in Europe, \ac{DSM} is projected to decrease renewable energy curtailment by 15.5 TWh by 2030, yielding economic savings of €4.6 billion compared to a scenario without \ac{DSM}~\cite{DNV2022}. Furthermore, \ac{DSM} mitigates annual greenhouse gas emissions in Europe
by 37.5 million tonnes, reinforcing its role in the transition to a low-carbon energy system~\cite{DNV2022}. Besides emission reduction, \ac{DSM} can enhance grid reliability and lower infrastructure reinforcement costs~\cite{strbac2020role}. \\

Both \acp{EV} \cite{liu2024cross,xu2018planning,powell2022charging,zhang2024sustainable} and buildings \cite{liu2023power,jackson2021building,tina2022technical,kleinebrahm2023two} offer the potential to modulate their electricity demand to support grid operations \cite{callaway2010achieving}. They have fundamentally different mechanisms and operational characteristics. The flexibility from \ac{EV} stems from controlled charging schedules and, in some cases, the discharging of the batteries through \ac{V2G} interfaces \cite{zhang2024sustainable}. Their flexibility is usually short-term (e.g., within a day) and limited by user mobility needs and the state of charge of the battery \cite{xu2018planning}. Due to their inherent mobility, the flexibility potential of \ac{EV}s demonstrates spatiotemporal variants \cite{powell2022charging}, posing challenges for prediction and reliable integration into demand-side flexibility programs. On the other hand, buildings offer flexibility based on their thermal inertia through \ac{HVAC} systems \cite{tina2022technical} and other smart appliances, such as lighting, dishwashers, and washing machines \cite{callaway2010achieving,li2021energy}. Unlike flexibility from \ac{EV}s, the flexibility potential of buildings is stationary and often more predictable, which can be reliable over a longer duration (hours to days). This is due to regular occupant and weather patterns~\cite{mathieu2011quantifying}. Together with an onsite \ac{PV}-battery system, an \ac{EMS} has the capability for load shifting and shedding \cite{li2021energy}, and local energy balancing \cite{kleinebrahm2023two,liu2023power}. \\

Modern power grids can use \ac{DSM} as a source of additional flexibility to maintain efficiency, reliability, and resiliency ~\cite{IEA2023,Ostergaard2021Energy}. For example, a case study in the United Kingdom demonstrated that the highly effective frequency response from \ac{V2G}-connected \acp{EV} enables secure frequency regulation, reducing fossil fuel-based power generation by 8~TWh~\cite{o2022frequency}. To fully unlock the potential of demand-side flexibility, a paradigm shift in power system operation is crucial. This transformation involves integrating energy supply and demand while creating market structures that incentivize demand-side flexibility participation. Additionally, it requires expanding flexibility services across both distribution and transmission networks \cite{strbac2018market} and establishing equitable cost-allocation frameworks. Achieving these goals requires substantial regulatory and social, economical, and technological advancements that account for diverse stakeholder interests and heterogeneous technical constraints \cite{matamala2024cost}. \\

This Review explores the grand challenges of implementing \ac{DSM} programs in modern power grids with high renewable penetration. Particularly, it considers the impact of sector coupling through the electrification of mobility and building. We present a holistic approach, a socio-techno-economic power system, that realizes the full potential of demand-side flexibility—both to support the grid and benefit consumers — integrating consumer social acceptance, advancements in control technology, and the electricity market. Rather than treating these aspects in isolation, we emphasize their interdependence. Finally, this Review outlines the remaining challenges and presents a forward-looking roadmap for a holistic framework.

\section*{A socio-techno-economic system}

Early research on demand-side flexibility focused on the development of control technologies, beginning with direct load control in the 1960s~\cite{buckingham1965remote}. In the early stages, these control techniques were assessed by their social, technical, and economic impacts~\cite{Abdoo1982load, Delgado1985demand}. Large power utilities promoted \ac{DSM} by emphasizing economic savings to foster the acceptance of end-users~\cite{Davis1983economics}. In 1978, Schweppe conceptualized modern power systems with demand-side flexibility as automated systems and marketplaces with active and flexible electricity consumers~\cite{Schweppe1978power}. Over the past two decades, the advancements in Information and Communication Technology, smart metering~\cite{eu2012smartmetering}, and improved transparency to power system operations~\cite{eu2013regulation543} have empowered the engagement of consumers in system-wide flexibility, and their awareness of benefits. \\

Social acceptance of consumers is key to increasing the adoption of \ac{DSM} programs. For instance, household characteristics, EV ownership, and evolving opinions impact participation willingness in \ac{DSM} programs~\cite{Yilmaz2019who}. Hence, emerging concepts such as the \textit{social license to automate}~\cite{Adams2021social} integrate social acceptance with automation in power systems with demand-side flexibility. Studies further differentiate between less socially acceptable flexibility—such as shifting small household appliance usage (e.g., lighting, dishwashers, washers, and dryers) \cite{Rinaldi2022what}—and more widely accepted approaches, including EV charging and leveraging building thermal mass. While social acceptance was also considered in the early research on demand-side flexibility, new paradigms with more accessible customer feedback can provide granular information to incentivize flexibility services. Further research is necessary to systematically integrate social acceptance in the automation and energy market, which enhances the adoption of \ac{DSM} programs.  \\

With continuous electricity market liberalization in power systems ~\cite{Wilson2002architecture,paul2008lessons}, new applications of demand-side flexibility are emerging across various timescales~\cite{palensky2011demand}. These opportunities incentivize consumers to participate in multiple markets simultaneously with key flexibility services \cite{villar2018flexibility}, in a techno-economic context ~\cite{Megel2015scheduling, Wang2020integrated}. Despite growing consumer interests in the energy market, automated demand response continues to play a central role. This is due to its cost-effectiveness, reliability, and minimal intrusiveness for end-users~\cite{Karjalainen2013should}. \\

\begin{figure*}[h]
    \centering
    \includegraphics[width=1.0\columnwidth]{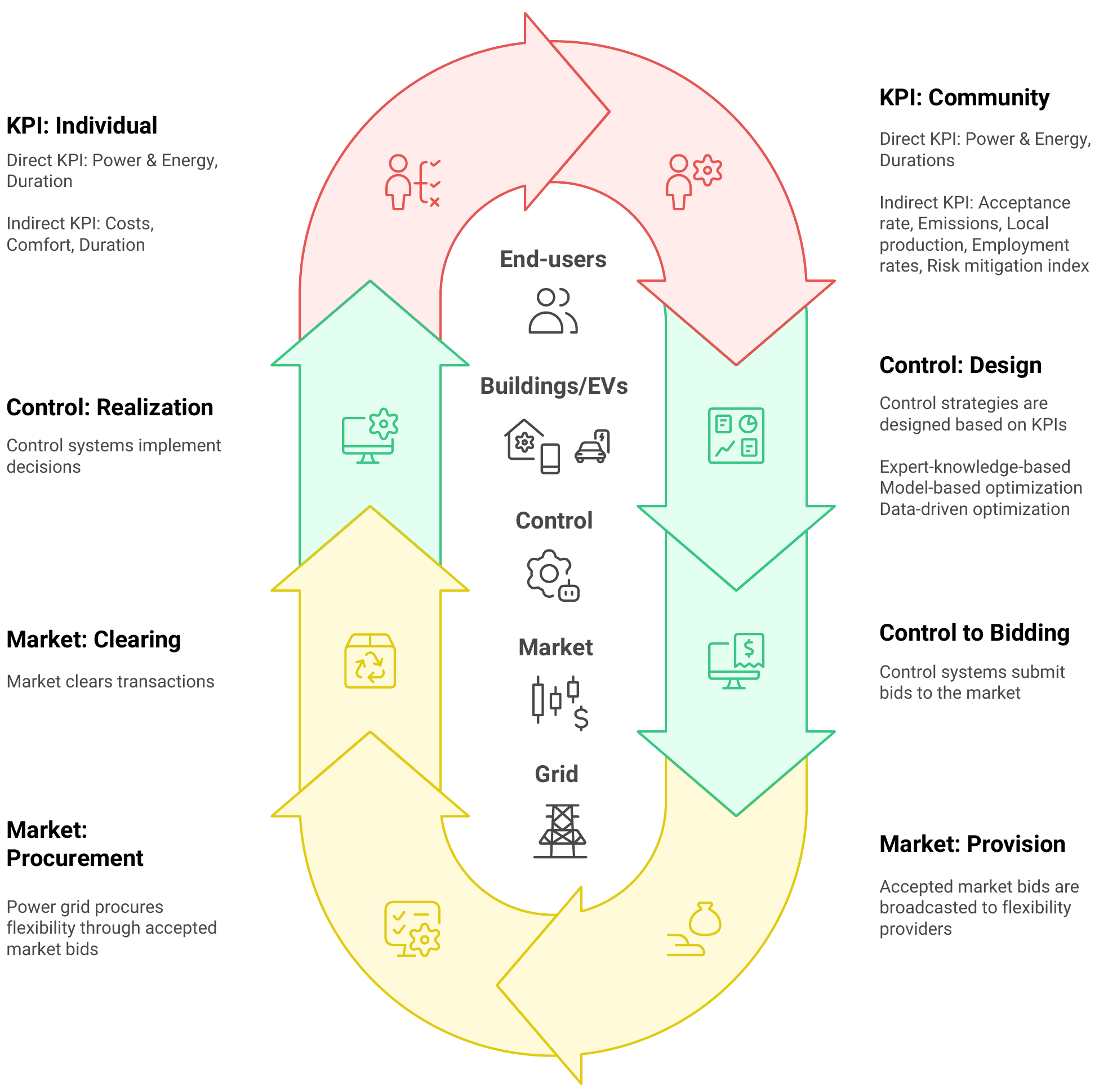} 
    \caption{\textbf{Overview of demand-side flexibility within a socio-techno-economic system.} The system is abstracted into five layers, from top to bottom: end-users, buildings/EVs, the control system, the market, and the power grid. The interlink between layers is illustrated through bi-directional information flows. Colored arrows highlight the discussion focuses of each section in this Review: \acp{KPI} (red), electricity market mechanisms (yellow), and control techniques (green). Starting from the end-users layer, demand-side flexibility quantification ensures users' comfort and aligns with social acceptance. By combining users' preferences with the physical system dynamics using flexibility \acp{KPI}, emerging control paradigms can explicitly incorporate the preference of flexibility providers into flexibility quantification. This quantification is then communicated to the market layer through bidding schemes. Using this information, markets match the offered flexibility with the demand while respecting the constraints of the end-users and the power grid. After market clearing, the accepted flexibility offers can be announced to individual flexibility providers. Finally, flexibility can be requested at the end-user level. This approach incorporates users' decisions, allowing them to actively contribute to the system's social welfare through behavioral adjustments~\cite{nagy2023ten}. }
    \label{fig:overall_structure}
\end{figure*}

In this review, we explore demand-side flexibility via a holistic approach, considering the end-users acceptance, electricity market mechanisms, and control techniques within a socio-techno-economic power system, which is depicted in \autoref{fig:overall_structure}. This framework highlights the closed-loop bidirectional flow of information between layers and the decision-making process from demand-side flexibility quantification to provision. However, it requires interactions between multiple stakeholders whose individual responsibility is not clearly defined yet~\cite{Greening2010demand}. Achieving a deeper interlink needs socio-techno-economic enablers and regulation solutions to key barriers of stakeholder cooperation~\cite{good2017review}. To facilitate the understanding of these interactions, we employ a high-level classification for each layer throughout the Review. \\

Our exploration begins with flexibility quantification, measuring the flexibility availability of consumers, which are categorized into direct and indirect \acp{KPI}. We then look into the existing electricity markets, which facilitate the systems in a centralized or decentralized manner. At the center of this holistic approach, we identify three categorized control approaches: expert-knowledge-based control, model-based optimization, and data-driven optimization. Beyond classification, automation systems impact the end user's comfort and external stakeholders; the inherent control decision-making process must be transparent and interpretable \cite{Stenner2017willingness, Michellod2022building, Adams2021social}, which are mandated by regulatory frameworks (i.e., the EU Artificial Intelligence Act~\cite{eu2024AIA}). In the end, we provide researchers and practitioners across multiple disciplines with insights into socio-techno-economic power systems in demand-side flexibility quantification and provisions. These insights can help overcome the barriers and drive the successful implementation of \ac{DSM} programs in the building and electric mobility sectors ~\cite[p. 83]{iea2017digitalization}.

\clearpage
\newpage
\section*{Flexibility quantification}\label{sec:definition_kpi}

The definition of demand-side flexibility is fundamental to both control implementation and the electricity market. Most measures directly or indirectly relate to power or energy metrics by adjusting energy blocks within an operational window, followed by predefined rules or optimization objectives. The definition of energy flexibility varies depending on design choices and operational strategies~\cite{reynders_energy_2018}. These variations stem from factors such as power profile~\cite{HEWITT2012_HPandTESchallenges}, the energy infrastructure~\cite{LUO2022123}, an exclusive focus on the electricity sector~\cite{Oldewurtel2010_PeakElectricity}, the energy price~\cite{LEDREAU2016_FlexBuildings}, the building performances~\cite{dhulst_demand_2015} and the system interaction~\cite{Kim2016_grid}.  Despite these differences, all these applications converge on a common ground: the capacity of flexibility providers to adapt their energy supply and demand with specific constraints. In this Review, we stick to the following general definition~\cite{reynders_energy_2018}:\\

\begin{itemize}
    \item [] \textit{``Energy flexibility is the ability to adapt the energy profile without jeopardizing technical and comfort constraints".} \\
\end{itemize}
Precise and timely quantification of end-user flexibility provides electricity market participants and flexibility activation controllers with the necessary information to ensure reliable and cost-effective grid support.\\ 

Quantified metrics such as \acp{KPI} of flexibility are usually designed for particular control tasks and market mechanisms. Their use in literature~\cite{li2021energy} highlights the need to consider \acp{KPI} across multiple control layers considering market dynamics within an integrated framework as the one in ~\autoref{fig:overall_structure}. A broader scope, ranging from individual buildings to their interactions with the grid and users, is important for developing effective flexibility measures. \\ 


\acp{KPI} can be categorized into direct and indirect metrics, capturing both technical and economic objectives for control and electricity markets, as well as considerations on human comfort, social acceptance, and sustainability-driven decarbonization. Advancing demand-side flexibility calls for additional KPIs, bridging technical and economic objectives with social considerations.

\subsubsection*{Direct KPIs}

Direct KPIs capture the key performance of flexibility providers in terms of power, energy, and their commitment to time durations \cite{LUO2022123}. They specifically quantify the amount of power shift over time in response to the flexibility activations~\cite{hurtado_quantifying_2017}.  Such actions include up-regulation by decreasing power demand and down-regulation by increasing power demand. The capability of consumers responding to such directive signals depends on their power ramping rate, energy availability, and power capacity~\cite{hurtado_quantifying_2017}, which are influenced by their physical characteristics, operational states, and environmental conditions (e.g., ambient temperature). Consequently, estimating flexibility potential is an inherently complex task that often requires an in-depth analysis by flexibility providers~\cite{ARTECONI2019_FlexQuantification}. \\


A key distinction among direct \acp{KPI} is whether they require a baseline scenario (an operating condition without external intervention~\cite{LUO2022123}). For baseline-dependent metrics, both the baseline scenario and a flexible scenario must be calculated. Currently, there is a lack of consensus in defining baselines, challenging the comparison of studies and practical implementation~\cite{ziras2021baselines}. In contrast, baseline-independent metrics can enhance scalability and adaptability for real-time decision-making. However, a lack of reference scenarios raises the risk of misinterpretation of activation and unintended operations, where the consumer responses are hard to predict. \\


A flexibility KPI, which integrates multiple performance metrics simultaneously, can provide a more comprehensive assessment. Absolute flexibility (measuring the capacity) and relative flexibility (ratios)~\cite{tang2021energy} as dual flexibility indices can be used to provide multiple services in terms of response speed, duration, and directives. Flexibility can also be assessed through shiftable and curtailable loads~\cite{OTTESEN2016Bidding}. A flexibility function~\cite{junker2018characterizing} with a dynamic nature captures the extent to which an end-user can respond to the grid requests. The grid operators control the demand via penalty signals. In addition, among these measures of flexibility, response time and committed power are significant for power grids. On the other hand, recovery time and energy variation matter most to end-users, indicating the speed of comfort restoration and the impact of \ac{DR} on energy use. However, integrating multiple system performance indices into a single KPI requires significant computational resources and data, which imposes challenges on feasibility of real-time implementation.


\subsubsection*{Indirect KPIs}
Flexibility quantification extends beyond engineering. It needs a multifaceted framework~\cite{AnwarETal} that integrates environmental and social perspectives to ensure successful implementation. Aside from direct \acp{KPI}, indirect \acp{KPI} measure flexibility in terms of economic (cost reduction), environmental (CO$_2$ emission reduction), and social (end-users benefits) impacts. Flexibility can be measured in terms of operating costs~\cite{Batic2016229, Patteeuw201680, Jelic2021_Social}, but it could also be combined by measuring the minimization of CO$_2$ emissions. CO$_2$ emission reduction is a critical environmental \ac{KPI} for flexibility quantification in various contexts, from building to industrial sectors~\cite{co2_emissions_kpi,co2_emissions_kpi2}. Operating on flexibility associated CO$_2$ emission index addresses climate change mitigation, regulatory compliance, and health and environmental benefits, which makes it an essential \ac{KPI} for entities committed to sustainability. \\


Social \acp{KPI}~\cite{AnwarETal, Li2018_Social, Reis2019_Social, Powells2019_Social,  Aduda2016_Social, Geneidy2020_Social, Mckenna2018_Social, Bergaentzle2021_Social, Gleue2021_Social, Kim2020_Social, Michaelis2017_Social, Ruokamo2018_Social, Schott2021_Social, Reis2020_Social, Noucier2020_Social, Nursimulu2015_Social}, from affordability and social equity to community self-sufficiency and risk governance, enrich the scope of demand-side flexibility as an engineering challenge with holistic perspective that integrates social dimensions into its design and implementation. Measuring flexibility through social KPIs provides valuable insights into its societal implications and benefits.\\

Affordability~\cite{Li2018_Social, Reis2019_Social, Powells2019_Social}, as one of the most widely discussed social \acp{KPI}, provides insights into the economic impact on individual households. By assessing energy cost reductions, particularly for lower-income groups, this sort of indirect \acp{KPI} highlights the potential flexibility to reduce energy poverty and contribute to more equitable energy systems. \\ 

Social equity and welfare distribution among different entities~\cite{AnwarETal, Reis2019_Social, Powells2019_Social} have been central to discussions on demand-side flexibility programs. They discuss whether all societal groups, regardless of their socio-economic status, have equal opportunities to access to their benefits. These \acp{KPI} provide valuable benchmarks for evaluating the inclusiveness of demand-side flexibility measures and their potential to bridge social disparities within energy systems.\\



Consumer satisfaction~\cite{Reis2019_Social, Aduda2016_Social, Geneidy2020_Social, Mckenna2018_Social, Bergaentzle2021_Social} reflects the consumers' perceptions of demand-side flexibility programs; it is a key social \acp{KPI} in consumer-centric energy systems.  Their perception includes personal experience, the level of control over energy usage, the tangible benefits received, thermal comfort, and indoor air quality. High satisfaction levels indicate well-designed programs that are responsive to consumer needs and preferences.\\

Social acceptance~\cite{Geneidy2020_Social, Bergaentzle2021_Social, Gleue2021_Social, Kim2020_Social,  Michaelis2017_Social, Ruokamo2018_Social} expands the individual satisfaction to the community level. It measures the community support for demand-side initiatives and reflects public awareness, attitudes, and the perceived value of flexibility measures. Similarly, health and well-being~\cite{Li2018_Social, Schott2021_Social} comprise broader environmental and health-related impacts of demand-side flexibility. More efficient energy use reduces harmful emissions and improves air quality, which directly benefits public health and eventually enhances the community's overall well-being.\\

In addition, community self-sufficiency~\cite{Reis2020_Social} reflects the ability of local energy resources and demand-side management to sustain its energy needs. Job creation~\cite{Michaelis2017_Social, Reis2020_Social, Noucier2020_Social} indicates the economic benefits of demand-side flexibility,  contributing to employment growth and skill development in the sustainable energy sector. Risk governance~\cite{Nursimulu2015_Social} ensures that robust frameworks for risk management support demand-side flexibility. It assesses the strategies to anticipate and mitigate potential adverse outcomes, thereby safeguarding energy system reliability and supply security.\\

We summarize the currently available direct and indirect KPIs in Supplementary Table 1. Despite increasing research efforts and discussion, there is no consensus on standardized flexibility \acp{KPI}. Most studies focus on direct \acp{KPI}, leveraging the measures of power, energy and committed duration. Nonetheless, indirect \acp{KPI} remain underexplored.   Notably, very few social \acp{KPI} are utilized in research studies on demand-side flexibility. Therefore, we recommend integrating social \acp{KPI} into the holistic assessment of control strategies and electricity market mechanisms for demand-side flexibility quantification and provision.



\section*{Electricity market mechanisms}\label{sec:market_mechanism}
Flexibility markets serve as platforms where flexibility providers (e.g., households, aggregators, distribution systems, and transmission systems) trade available flexible energy from their assets in response to grid needs \cite{jin2020local,eid2016managing}. The amount of flexible energy available for bidding in the markets is determined by the adopted control strategies and \acp{KPI}, as shown in ~\autoref{fig:overall_structure}. Effective market designs can enhance demand-side flexibility dispatch by integrating both spatial and temporal dimensions~\cite{huo2020spatio}. The spatial dimension accounts for the physical locations where demand-side flexibility resources are available or needed subject to grid constraints. Meanwhile, the temporal dimension considers variations in energy demand and availability over time. By incorporating these factors, flexibility markets can effectively balance supply and demand through short- and long-term energy trading or dynamic pricing mechanisms, where energy costs vary based on the time of use~\cite{Hussain2023flexibility}.\\

Electricity markets are typically categorized into centralized and decentralized frameworks. In the centralized approach, a central operator directly communicates with and controls all flexible \acp{DER}, aiming to optimize multiple objectives. However, this approach requires significant information sharing, which raises concerns about privacy and trust among end-users and other stakeholders. In particular, aggregators, mostly privately owned companies, prefer to keep their operations confidential to maintain a competitive advantage~\cite{TKACHUK2023101146}. The requirement to surrender full control over flexible resources, combined with concerns about privacy and trust, results in a low willingness among end-users to provide flexibility~\cite{morstyn2019}. Conversely, decentralized approaches enable end-users or aggregators to engage directly in energy transactions without relying on a single operator, thus preserving their privacy and maximizing individual benefits~\cite{TKACHUK2023101146}. However, this decentralization may come at the cost of not achieving a global optimum.\\

In this Review, we extend the description of centralized and decentralized market frameworks by incorporating a comparison of their adoption within the holistic framework based on different control strategies and \acp{KPI}. Drawing from state-of-the-art research and existing pilot projects, we examine various market mechanisms and the extent to which they have been adopted in practice.
 
\begin{tcolorbox}[title = \themybox: Flexibility services, colframe=nature_red, colback=nature_gray, coltitle=white, fonttitle=\bfseries, sharp corners, boxrule=1pt, width=\textwidth]
\refstepcounter{mybox}\label{box:def_flexibility}
To effectively implement centralized and decentralized market frameworks, flexibility services have been developed to enhance grid stability and efficiency. We summarize key flexibility services identified in the literature and existing pilot projects:

\begin{itemize}
\item \textit{Peak shaving/shifting} aims at reducing or shifting electricity consumption during peak demand periods, typically achieved by implementing energy management systems that respond to price signals~\cite{lampropoulos2019framework}. 

\item \textit{Demand regulation} involves actively managing and controlling electricity demand to align with available supply, commonly achieved through energy management systems, similar to peak shaving/shifting~\cite{Plaum2022}. 

\item \textit{Congestion management} aims to alleviate congestion in electricity transmission and distribution networks, facilitated by 
the availability of local markets or spatio-temporally varying pricing mechanisms~\cite{HAQUE201765}. 

\item \textit{Frequency reserve} involves maintaining the balance between electricity supply and demand to keep the system frequency within acceptable limits, facilitated by the availability of ancillary service markets~\cite{MOTALLEB2016439}.
\end{itemize}
\end{tcolorbox}

\subsection*{Centralized markets}
Two key barriers hinder full access to the demand-side flexibility in a centralized market-based procurement model, relying on full trust from end-users and lack of privacy concerns~\cite{distribution_systems_working_group_ceer_2020}: flexibility service pricing and power dispatching.\\

\acp{LFM} facilitate flexibility trading between a \ac{DSO} and providers, such as aggregators~\cite{liu2021distribution}, to optimize service pricing. However, their location dependency limits market liquidity~\cite{grvzanic2021review}. A higher market liquidity has many flexibility buyers and providers to have quick trades at stable prices. Limited liquidity can result in price volatility and inefficiencies in matching flexibility supply and demand. Pilot projects further reveal that short-term offers on \acp{LFM} do not support long-term grid planning. The baseline definitions of flexibility services remain prone to manipulation and/or lack transparency, challenging accurate pricing~\cite{valarezo2021, SCHITTEKATTE2020, ceer2018flexibility, ziras2021baselines}.\\

To address these challenges, aggregator-activated demand response schemes have emerged as a complementary solution to unlock demand-side flexibility. In these schemes, energy retailers adjust aggregated demand in response to tariffs while enabling energy trading~\cite{mohseni2021modelling}. Optimization-based methods further manage uncertainties and maximize flexibility profits using shiftable and curtailable loads~\cite{Lu2020aggregator, tang2022, OTTESEN2016Bidding}. Similarly, game-theoretic approaches model strategic interactions between market players, optimizing flexibility prices through intermediary aggregators~\cite{mohseni2021modelling, attar2022congestion}.
In these approaches, the interactions between \acp{DSO} and aggregators can be modeled as a Stackelberg game, where either the \acp{DSO} act as leaders and aggregators as followers~\cite{mohseni2021modelling, amasyali2022data, fernandez2018game}, or vice versa~\cite{Huang2020game, Nasiris2020stackelberg}. Alternatively, locational marginal prices for demand response can be used~\cite{Chandra2021, wang2023coordinated}, or transactive prices can be derived based on internal operation strategies to ensure cost recovery~\cite{Wang2020pricing}.\\

Optimally dispatching flexibility requires consideration of multiple factors, such as sensitive information sharing, grid constraints, and fairness, which can become computationally complex~\cite{kirschen2018fundamentals}.
Centralized load management strategies, often involving a central controller, optimize energy dispatch across multiple units~\cite{martirano2013, Agheb2018}. However, as the number of units increases, scalability becomes a challenge. To address this, distributed and meta-heuristic approaches enable decentralized decision-making, improving privacy and scalability but increasing communication overhead~\cite{luo2017rolling, lampropoulos2019framework}. In addition, game-theoretic approaches facilitate autonomous optimization of flexible device consumption, balancing efficiency, privacy, and computational feasibility in large-scale energy systems~\cite{tang2022model}.

\subsection*{Decentralized markets}
Decentralized markets enable local energy trading among end-users and energy hubs, leveraging \ac{DER} flexibility to shift demand, enhance transparency, and promote privacy while reducing fraud. By fostering a stronger sense of community, these markets maximize both individual and societal benefits~\cite{SOTO2021116268}. \\

Local energy markets, derived by price-based optimization and transactive energy designs, facilitate dynamic energy exchanges and grid services among diverse participants~\cite{moret2018energy}. However, they can expose small consumers to price fluctuations~\cite{joskow2000we}. To address this, \ac{P2P} trading has emerged, allowing smaller consumers to participate and offer flexibility services~\cite{vahidinasab2021participation}. Various market structures, including community-centric markets~\cite{saif2022impact}, game-theoretic-based pricing models~\cite{Bahrami2019potencialgame}, and auction-based mechanisms~\cite{Khorasany2021doubleauction, Khorasany2019doubleauction} enhance participation but require cooperation for convergence. Participants must incorporate penalty terms rather than solely maximizing individual utility. In addition, new mechanisms are needed to integrate \acp{DSO} into \ac{P2P} trading, enabling them to negotiate flexibility procurement with aggregators and end-users~\cite{morstyn2018using}. In response, decentralized market designs have been proposed, allowing \acp{DSO} to manage local demand constraints by procuring flexibility from competing aggregators. These aggregators, in turn, incentivize end-users to participate~\cite{morstyn2019}.\\


A summary of the flexibility market mechanisms is presented in Supplementary Table 2. Centralized markets offer structured flexibility management but raise concerns over privacy, trust, and participation barriers. In contrast, decentralized markets provide greater autonomy and transparency yet face challenges in fairness and coordination. Within the socio-techno-economic system, control strategies vary across market types: model-based optimization dominates in \acp{LFM} and tariff systems, while data-driven approaches are emerging in \ac{P2P} markets. Traditional expert-knowledge-based methods are becoming obsolete as they do not account for evolving market signals. Notably, flexibility services are essential for grid decarbonization but face challenges in pricing and power dispatch. At present, there is no consensus on the preferred flexibility market mechanism. In addition, the diverse market frameworks create difficulties in comparison and hinder efficient information exchange across system layers in~\autoref{fig:overall_structure}, limiting the integration within a broader socio-techno-economic system.

\section*{Enabling flexibility provision}\label{sec:enabling_strategy}
\begin{figure*}[htbp]
    \centering
    \includegraphics[width=0.85\columnwidth]{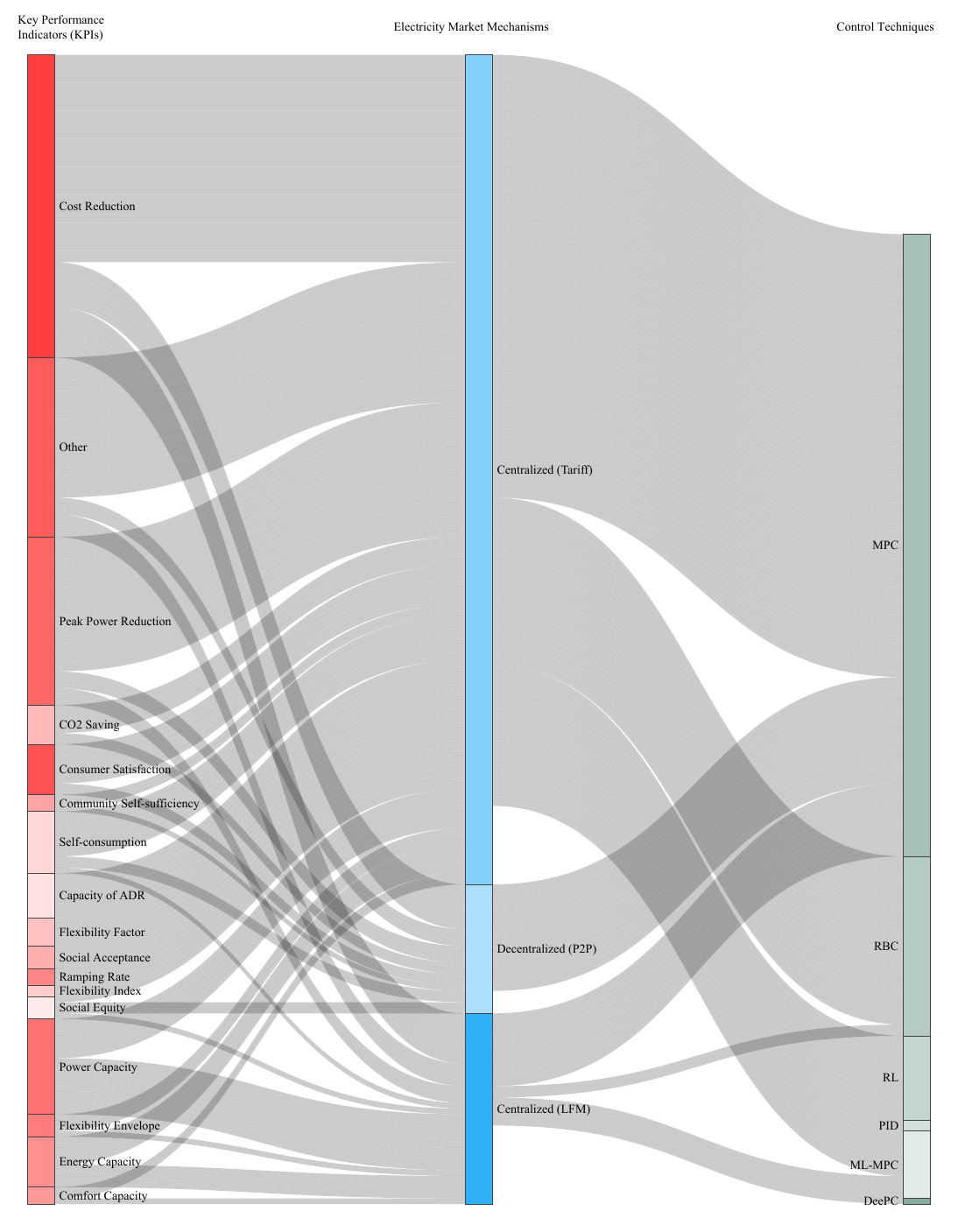}
    \caption{\textbf{Overview of the interconnections among flexibility KPIs, market mechanisms, and control techniques.} \ac{RBC} and \ac{PID} remain industrial state-of-the-art control techniques. They both have been mainly investigated in tariff-based markets with limited use cases involving market bidding. Due to the inability of these two approaches to handle complex decision spaces, they have rarely been explored in more sophisticated market structures, such as \ac{P2P} trading. In addition, implementation with \ac{RBC} for signal tracking is rarely reported. Given its reliance on expert knowledge and a predefined space of control policies, \ac{RBC}'s performance is commonly expected to be lower than other optimization-based techniques, which makes them a benchmark controller in various studies. In contrast, optimization-based approaches such as \ac{MPC}~\cite{henze2005experimental,oldewurtel2010energy,oldewurtel2012use}, \ac{ML}-based \ac{MPC}~\cite{aswani2012reducing}, and \ac{DeePC}~\cite{coulson2019data,o2022data,Lian2023adaptive,Yin2024data}, and 
    \ac{RL}~\cite{Gregor2003Evaluation,ruelens2016residential} are expected to deliver a good performance if the underlying physical process is sufficiently captured and the optimal control problem can be solved promptly. There is a clearer preference for \ac{MPC} over other techniques, and it has been applied to all market mechanisms.  This is attributed to its strong performance in automated decision-making within complex environments and its early appearance compared to \ac{RL}, \ac{ML}-based \ac{MPC}, and \ac{DeePC}. 
    Nonetheless, the requirement for an accurate system model often limits the performance of \ac{MPC}, where data-driven approaches offer a promising alternative. Notably, \ac{ML}-based \ac{MPC} addresses the scalability limits by directly replacing first-principle modeling with \ac{ML}. It has been explored in tariff- and bidding-based market frameworks. Meanwhile, \ac{RL} has primarily found its applications in tariff-based markets but not other market mechanisms. Note that we visualize popular \acp{KPI} in the literature, including cost reduction, peak power reduction, and self-consumption. The discrepancy between flows from the left and right is because existing studies do not always fully consider flexibility KPIs, market mechanisms, or control techniques. The segment labeled “Other” in the left bar contains additional \acp{KPI}, including comfort recovery, flexibility function, flexibility storage, flexibility revenue, and public health.}
    \label{fig:summarySankey}
\end{figure*}

The flexibility quantification and provisions from the demand side can be divided into three key components: (1) flexibility \acp{KPI}, (2) market mechanisms, and (3) control techniques. Control techniques play a central role in adjusting flexibility resources guided by the measures of flexibility \acp{KPI} and outcomes of market mechanisms. We visualize the interconnections among these three components in \autoref{fig:summarySankey}. \\

\begin{figure*}[htbp]
    \centering
    \includegraphics[width=1\columnwidth]{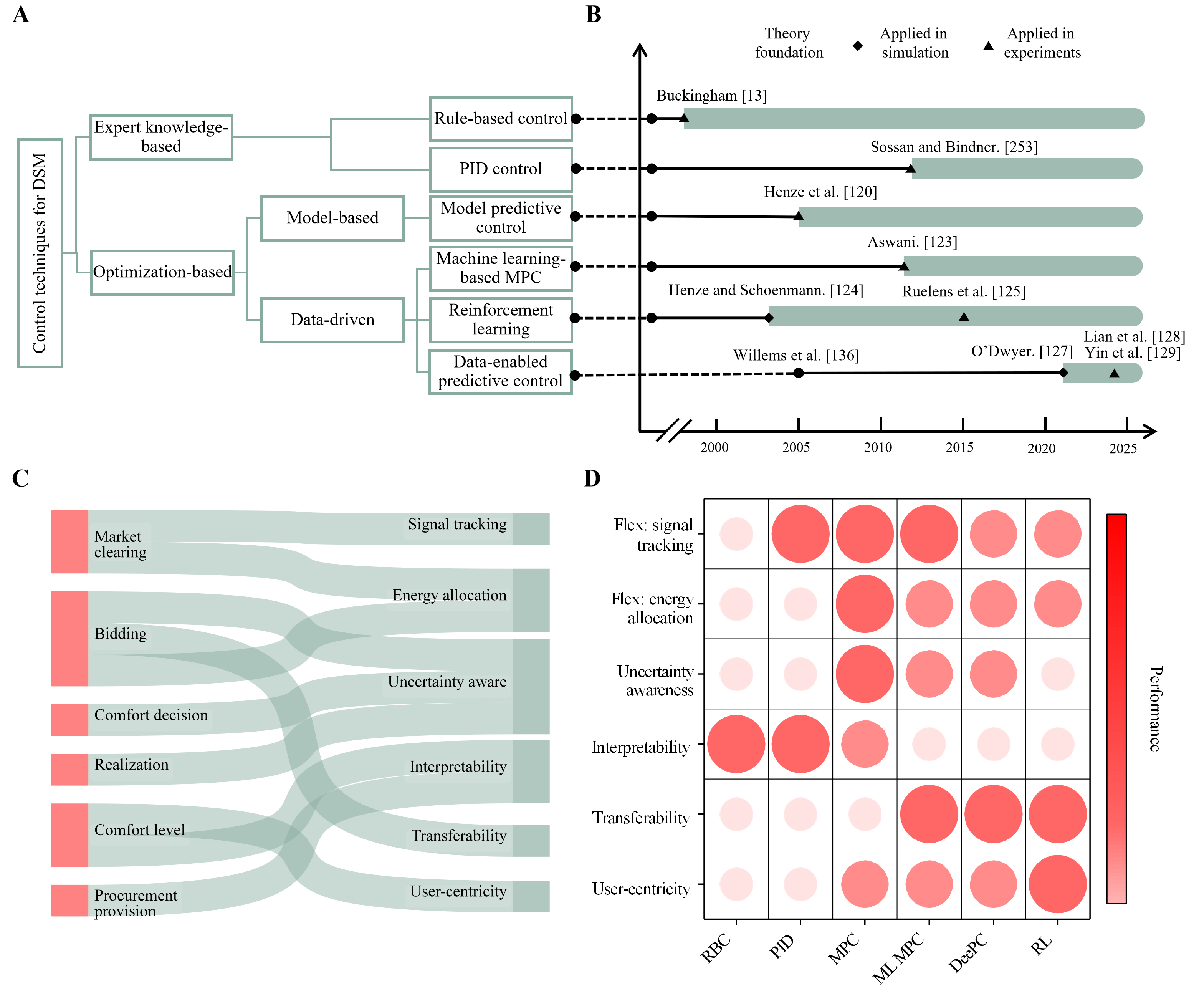}
    \caption{\textbf{Control techniques with applications to demand-side flexibility.} (\textbf{A}) Classification of control techniques applicable to demand-side flexibility, divided into expert knowledge-based and optimization-based categories.  (\textbf{B}) Timeline highlighting the progression of classified control techniques from theoretical foundations through simulation studies to experimental applications, with selected contributions. (\textbf{C}) A Sankey diagram highlights the interlinks between the steps of flexibility provisions (on the left side), as also shown in \autoref{fig:overall_structure} and the desired control performance (on the right side). (\textbf{D}) Performance comparison of control techniques. The size and color of the solid circles qualitatively represent the empirical performance ratings of control techniques, where larger circles and darker colors indicate higher performance ratings. Additionally, the energy system-related services are categorized into signal tracking and energy allocation. While signal tracking is relevant for frequency reserve, energy allocation covers the suitability of tasks like peak shaving, congestion management, and demand regulation. Note: RBC, rule-based control; PID, proportional-integral–derivative; MPC, model predictive control; ML-MPC, machine learning-based model predictive control; DeePC, data-enabled predictive control; RL, reinforcement learning.}
    \label{fig:control_overview}
\end{figure*}

The current control schemes can be broadly categorized based on whether the controller design relies on expert knowledge or optimization techniques. The expert-knowledge-based controllers include \ac{RBC} and \ac{PID} control strategies. The optimization-based controllers consist of both model-based and data-driven techniques. Data-driven techniques are mainly motivated by the need to address the scalability limits of model-based approaches, which require extensive details of the underlying system dynamics and a considerable amount of engineering effort. This categorization, along with a timeline highlighting the key milestones in the historical development of each technique, is shown in \autoref{fig:control_overview}(a) and \autoref{fig:control_overview}(b). The mathematical formulations of control techniques have been summarized in Supplementary Note 1. 


\subsection*{Existing control techniques}\label{subsec:enabling_strategy}

\textit{Rule-based controller:} \acp{RBC} are broadly used in industry and often serve as benchmark control schemes~\cite{Pinto2021RL,Lian2023adaptive}. However, their limitations are in ensuring optimal control decisions and handling uncertainties. Significant engineering efforts are needed to define the rule set for complex tasks. Most importantly, the quantification and maximization of the flexibility \acp{KPI} can hardly be achieved via \ac{RBC} due to its simplistic and static nature.\\

\textit{PID controller:} The \ac{PID} controller can be a versatile and robust solution in many applications when properly tuned, ranging from simple home appliances to complex industrial processes~\cite{Kurz1978Development,Schumann1982Digital}. Additionally, it can provide precise and smooth control setpoints once the parameters are well-tuned. Achieving this desired performance requires careful adjustment of its three parameters to ensure the expected response, stability, and fast reaction without excessive oscillation or steady-state error. However, \ac{PID} controllers ignore future knowledge and cannot easily incorporate operational constraints. Although the PID controller stands out for its performance and broad adoption for signal tracking, it has not found many applications in energy allocation tasks. \\

\textit{Model predictive control (MPC):} \ac{MPC} is an advanced control strategy that optimizes system behavior over a future time horizon by solving a constrained optimization problem at each time step. It uses a dynamic model of the system to predict future states and iteratively updates control decisions based on real-time measurements and forecasts \cite{kaya1978modeling,garcia1989model,henze2005experimental}. The applications of \ac{MPC} in building control have been extensively reviewed in~\cite{Drgona2020all}. \ac{MPC} framework can incorporate user preferences as constraints via recursive optimization. \ac{MPC} can handle uncertainty by leveraging probabilistic information and utilizing stochastic optimization techniques. It also has been extended to multi-agent settings with distributed \ac{MPC}~\cite{Corbin2017dmpc}, and hierarchical multi-level control schemer~\cite{xu2020hierarchical}. The importance of \ac{MPC} in enhancing power grid reliability and improving energy efficiency has also been demonstrated with building operations~\cite{Zhang2016}. \\

\textit{Machine learning-based MPC:} \ac{ML}-based \ac{MPC} controller has been utilized in building \ac{HVAC} systems by leveraging deep neural networks to capture the complex dynamics of physical processes \cite{curtiss1993adaptive,aswani2012reducing}, such as natural ventilation~\cite{chen2020transfer}. To reduce the dependency on extensive data collection when applied to new buildings, only a few parameters are re-trained. At the same time, most pre-trained deep neural network model layers are retained.\\

\textit{Data-enabled predictive control (DeePC):} Based on the fundamental lemma~\cite{willems2005note}, several variants of data-enabled predictive control have been proposed to address uncertainties in practical implementations~\cite{huang2023robust,Yin2024data,shi2025adaptive}. Multiple regularization techniques have been shown effective~\cite{huang2023robust} to account for nonlinear system dynamics and noises. In addition, uncertainty measures can be incorporated through chance-constrained formulation in a stochastic data-enabled predictive control framework~\cite{Yin2024data}, with experimental evaluations applied to demand-side flexibility tasks. At present, few studies have applied data-enabled predictive control for demand-side flexibility provision~\cite{shi2025adaptive}.\\

\textit{Reinforcement learning (RL):} Reinforcement learning started being explored in controlling thermal loads~\cite{Gregor2003Evaluation,ruelens2016reinforcement}. In the context of the demand-flexibility provision, \ac{RL} can learn from the experience for (1) improving energy efficiency while maintaining occupant comfort~\cite{Pinto2021RL,Lissa2021RL}, (2) enabling demand response programs~\cite{Peirelinck2018RL,VAZQUEZCANTELI20191072}, (3) energy storage system management~\cite{Zhou2022MLRL}, (4) integrating renewable energy sources~\cite{Zhou2022ML,VAZQUEZCANTELI20191072}, and (5) \ac{EV} charging scheduling~\cite{VAZQUEZCANTELI20191072}. \\

A summary of control technique applications, including the types of control techniques, flexibility services, and a brief description of each study,  is provided in Supplementary Table 3.

\subsection*{Operational uncertainty}\label{sec:uncertainty}

Considering uncertainty is crucial when quantifying and providing demand-design flexibility. These uncertainties usually stem from user behavior~\cite{martinez2022}, load and renewable generations~\cite{OTTESEN2016Bidding}, load rebound effects~\cite{Mai2015mpc}, and electricity prices~\cite{Agheb2018}. Deterministic control approaches can potentially lead to a mismatch between the quantified flexibility and availability during real-time operation. Given the safety-critical nature of power systems, carefully addressing operational uncertainty is important to maintain system reliability and avoid market penalties due to unfulfilled flexibility commitments. \\


Multi-layer hierarchical frameworks can be a promising solution for flexibility provision to address uncertainties at different levels~\cite{Agheb2018}. The first layer addresses robust energy allocation between the day-ahead and reserve markets, considering day-ahead price uncertainty. The second layer employs a bi-level optimization to determine optimal aggregator bidding strategies in the market. Finally, the third layer deals with the optimal dispatch of flexible resources, considering a performance factor to evaluate each resource based on the reliability of its past responses. \\
 


Studies that explicitly leverage stochastic or robust optimization frameworks to account for uncertainty mainly employ \ac{MPC} or \ac{DeePC} (Supplementary Table 3). Other techniques employ deterministic equivalence and do not explicitly account for uncertainty. This is attributed to the comparably seamless way of incorporating uncertainty in the optimization-based frameworks. The stochastic framework can account for epistemic uncertainty due to inaccurate models and forecast errors. In contrast, aleatory uncertainties refer to the fundamental randomness that cannot be incorporated.


\begin{tcolorbox}[title = \themybox: Interpretability of Controllers, colframe=nature_red, colback=nature_gray, coltitle=white, fonttitle=\bfseries, sharp corners, boxrule=1pt, width=\textwidth]
\refstepcounter{mybox}\label{box:def_flexibility}
The European \ac{AI} Act \cite[Article 13]{eu2024AIA} mandates that a ``\textit{high-risk \ac{AI} system shall be designed and developed in such a way as to ensure that their operation is sufficiently transparent to enable deployers to interpret a system's output and use it appropriately}" while adopting a framework based on risk levels. The high-risk classification of the energy sector due to potentially severe negative impacts on well-being in case of electric power grid failures~\cite{eu2024AIA,heymann2023operating} highlights the criticality of interpretable decision processes in demand-side flexibility. 

\end{tcolorbox}

\subsection*{Decision Interpretability}\label{sec:definitions}

Even though model interpretability for power and building energy systems has been discussed~\cite{machlev2022interp, chen2023interp}, the interpretable rule extraction into control techniques for traceable decisions and building user trust remains limited. Due to little research on interpretable controllers, we draw on general definitions of interpretable \ac{ML} models, along with their challenges and existing tools, to propose a framework incorporating interpretability into control techniques. \\

Studies in interpretability have mainly focused on \ac{ML} models, and the concept can be categorized into inherently interpretable and \ac{XAI}~\cite{rudin2022interpretable}. Inherently interpretable models are designed to be self-explanatory and respect domain-specific constraints, making their outputs easier to understand and diagnose.
In contrast, \ac{XAI} is a model-agnostic approach to explain non-inherently interpretable models using post-hoc methods~\cite{rudin2022interpretable}. Such approaches, \ac{SHAP} and \ac{LIME}, can be used to explain their outputs through approximation models. 
For example, black-box models like \ac{ANN} can capture non-linear relationships better than naturally interpretable models but lack inherent interpretability, which requires additional explanation methods to make their decisions understandable. \\


So far, there is no standard set of \acp{KPI} for quantifying interpretability, and this area remains underexplored~\cite{machlev2022interp}. Evaluation is one of the most essential aspects of interpretable \ac{ML}~\cite{murphy2023}, but it varies across application domains. Therefore, the \acp{KPI} should be tailored to the specific needs of each domain while ensuring consensus among stakeholders~\cite{rudin2022interpretable, murphy2023}. \\


The development of inherently interpretable \ac{ML} models is still in its early stages ~\cite{chen2023interp,machlev2022interp,molnar2020pitfalls}. Although control systems increasingly rely on complex optimization problems that are not inherently interpretable, research on improving the interpretability of control law is still limited. Previous efforts have focused on scalability, which uses surrogate models to approximate computationally expensive optimization problems~\cite{drgovna2018rule_ex}.\\
Future research on demand-side flexibility can benefit from emphasizing inherently interpretable \ac{ML} models to align with regulatory frameworks. Explained black box models have many uncertainties, especially problematic for high-stake decisions. The application of surrogate models, particularly simpler ones like decision trees, to improve the scalability and interpretability of advanced control algorithms remains an emerging topic. \\

\begin{figure}
    \centering
    \includegraphics[width=0.8\columnwidth]{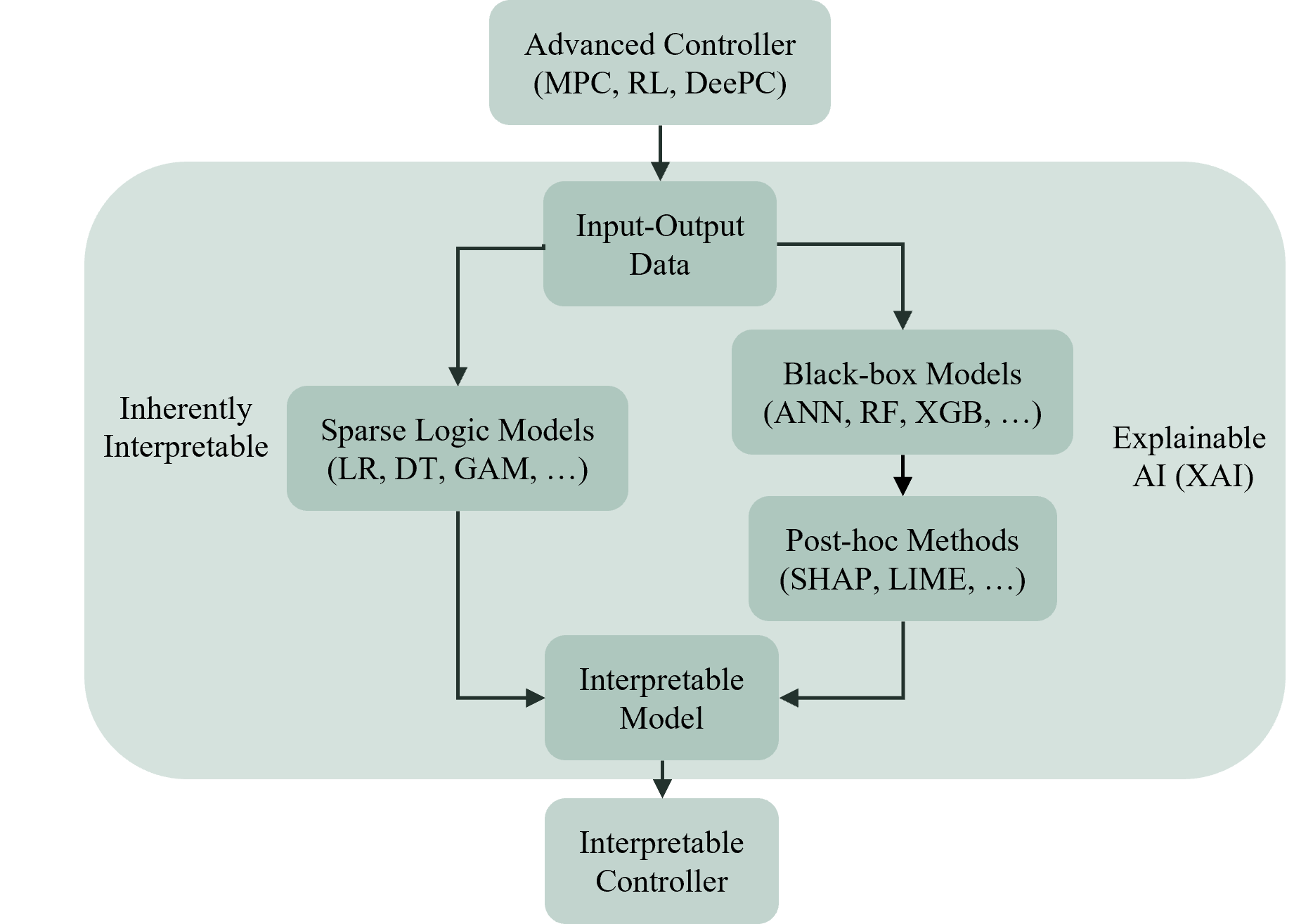}
    \caption{\textbf{A conceptual framework to enable interpretability across control techniques, inspired by the latest developments in interpretable machine learning.} From a system's perspective, a controller functions similarly to a physical process, which maps inputs to outputs. However, the interpretability becomes unclear if this mapping describes complex functions. Two promising approaches in interpretable machine learning can enhance the transparency of complex control algorithms as a posterior module. The first approach involves using inherently interpretable techniques, such as LR, which offer insights into the weights of input features and allow us to trace decisions back to the inputs. The second approach combines a more complex, black-box model, like an ANN, with a post-hoc method, such as LIME, to help us understand the controller decisions.
    Note: ANN, artificial neural networks; DeePC, data-enabled predictive control; DT, decision tree; GAM, generalized additive model; LIME, local interpretable model-agnostic explanations; LR, linear regression; MPC, model predictive control; RF, random forest; RL, reinforcement learning; SHAP, shapley additive explanations; XAI, explainable artificial intelligence; XGB, extreme gradient boosting.}
    \label{fig:interpretability}
\end{figure}

While \ac{RBC} excels in simplicity and interpretability, the standard \ac{RBC} only reacts to the most recent feedback and does not leverage future information. In contrast, other control techniques suffer from interpretability difficulties due to their reliance on optimization-based decision-making or the inherently exploratory nature of \ac{RL}. Nonetheless, \ac{MPC} remains advantageous over other control techniques since it uses first-principle modeling. Note that black-box modeling commonly lacks physical consistency. The interpretability framework explained in \autoref{fig:interpretability} can facilitate the understanding of the decision-making processes of stakeholders, including control engineers, system operators, end-users, and policymakers. This understanding is crucial for identifying model limitations and security gaps to ensure process safety and reliability, building trust in applications, and promoting acceptance of solutions.

\subsection*{Transferability of Controllers}
\label{subsec: Transferability}
For effective and widespread adoption across potentially billions of distributed devices~\cite[p. 83]{iea2017digitalization}, control algorithms must demonstrate high performance with interpretability while requesting minimal manual engineering when applied to various devices.  This brings us to the concept of transferability. A transferrable algorithm can be applied to similar problems with minimal additional engineering effort \cite{weiss2016survey}. For example, as defined in~\cite{pinto2022transfer}, transferability implies that an algorithm designed to control the temperature in one zone can be used to control the temperature in another zone within the same building or in a different building, delivering optimal performance either immediately or after a brief adaptation period. \\

Most research on the transferability of control algorithms in the building and electric mobility sectors has focused on \ac{RL} and transfer learning. The application of transfer learning in smart buildings is reviewed in \cite{pinto2022transfer}, covering aspects such as building control, load prediction, building dynamic models, and occupancy. It highlights the role of deep learning in transfer learning applications and emphasizes its potential to enhance building energy management strategies. Furthermore, transfer learning has been applied in \ac{DR} for decarbonizing the energy sector, and most research has focused primarily on transfer learning in electricity demand prediction~\cite{peirelinck2022transfer}. However,  there is increasing attention to transferring the controllers, particularly within building control systems using \ac{RL} techniques. \\
The transferability of controllers has been predominantly explored by \ac{RL} to address the scalability challenges due to intensive pre-deployment training. Additionally, several studies provide experimental investigation in the case of \ac{DeePC} \cite{Yin2024data}. The user-centric feature requires adaptiveness to evolving user preferences and social \acp{KPI}, as indicated in Supplementary Table 1. While \ac{RL} inherently embeds this adaptiveness in its methodology, other optimization-based techniques, such as \ac{MPC}, \ac{ML}-based \ac{MPC}, and \ac{DeePC} can incorporate end-user's preferences by adjusting the constraints of optimal control problems. Ad-hoc solutions such as regular updates of identified models or data-driven presentations can be employed to adapt to the evolving system dynamics over time \cite{Lian2023adaptive}. 
Furthermore, hybrid strategies that combine multiple control techniques have been proposed~\cite{langer2020,ZHOU2023106536,drgovna2018approximate,yang2021experiment}. This review can be a foundation for developing such hybrid approaches, enabling the combination of desired features or supporting context-aware controller design. A comparison of different control techniques is summarized in~\autoref{fig:control_overview}(c) and~\autoref{fig:control_overview}(d) .

\section*{Challenges and outlooks}\label{sec:discussion}
Although demand-side flexibility shows remarkable potential in power and control engineering, electricity markets, and social sciences, there remain opportunities to further fully harness their potential. Field implementation faces hurdles on societal, technical, and economic fronts, which slow adoption despite its promise. Addressing these challenges and their interactions also paves the way for facilitating the deployment of \ac{DSM} programs in practice, as discussed here.


\subsection*{Societal aspects}
Energy end-users prioritize comfort-related objectives, such as maintaining room temperatures or adequate lighting, over operational robustness and cost-effectiveness. End-users consider electricity as a necessity rather than just a regular commodity, making its value greater than what its market price reflects. This indispensable nature implies that consumers are not easily motivated to change their usage patterns just for financial savings. The benefit of participating in \ac{DSM} may be relatively small compared to the perceived inconvenience. Conversely, discomfort must be substantially remunerated to ensure acceptance~\cite{VAZQUEZCANTELI20191072}. \\

Moreover, a significant mismatch often exists between flexibility providers and resource owners. For example, tenants may lack the authority to implement flexible operations that require modifications to the property. On the other hand, property owners may not perceive any direct benefits, as flexibility needs and comfort constraints primarily affect tenants. This misalignment limits the exploitation of demand-side flexibility, even when incentive schemes are available~\cite{sorrell2000reducing}. \\

Ensuring operational flexibility is most efficient when integrated during building conceptualization and planning stages. The low retrofit rate, around 1\% per year in most European countries~\cite{cordis2021retrofitting}, hinders the large-scale implementation of energy-flexible buildings. Building owners and architects prioritize architectural and functional requirements, and often overlook future operational stage flexibility~\cite{planning_neglect}. Including it later for market participation increases construction costs, which makes it a less attractive option. \\

Another significant barrier is due to the heterogeneity of flexibility \acp{KPI}. The diverse flexibility \acp{KPI} introduces a notable barrier to energy flexibility literacy. We hypothesize that this ambiguity and complexity will reduce end users' trust, which is key for solutions acceptance~\cite{Stenner2017willingness}. End-users' comfort and preferences significantly affect the potential of demand-side flexibility. It is important to address various dimensions of comfort adaptation, including behavioral, physiological, and psychological aspects~\cite{deDear1998}. Thermal comfort with temperature limits can be explicitly encoded in \ac{RBC} as situations for the decision, in 
\ac{PID} as a tracked reference, in \ac{MPC}, \ac{ML}-based \ac{MPC} and \ac{DeePC} via explicit constraint satisfaction and \ac{RL} as changes on the reward of interaction. However, accommodating further dimensions of behavioral, physiological, and psychological factors poses a greater challenge, as they cannot be directly measured and embedded into the technical setups. A key step in accelerating the adoption of demand-side flexibility is developing a technology-agnostic \ac{KPI} with reduced complexity. Such simplification is important to raise awareness and promote energy literacy \cite{VandenBroek2019}. Addressing standardization gaps and establishing evaluation \acp{KPI} for interpretability and transferability is also essential. It requires consensus among different stakeholders~\cite{rudin2022interpretable, murphy2023}. \\

\subsection*{Technical aspects}
In a traditional centralized energy system, energy flows unidirectionally from central production to consumers, adapting supply to meet demand. When grid capacities reach their limits, extensions are made to maintain reliability. While this approach ensures reliable grid operation, it is not cost-efficient. \ac{LFM} and new demand-side technologies provide more economical alternatives, yet traditional approaches persist, particularly when \ac{DSO} processes and competencies require further development~\cite{planning_neglect}.\\

The energy industry mostly relies on rule-based control for simple processes. A shift towards sector-coupled and optimization to data-driven approaches requires system planners and integrators to develop new competencies and processes to handle these new and arguably more complex approaches. Moreover, systems planners and integrators focus on competitive pricing, creating a low-margin environment where there is little room for the development of new skills and competencies beyond established and standard solutions~\cite{planning_neglect}.\\


Facility managers of buildings oversee the reliable operation of multiple buildings and systems with minimal maintenance. They also serve as the first responders in case of malfunctions or emergencies. In an emergency, it is important to have a traceable and consistent system behavior across all buildings. To facilitate diagnosis, standardized operational \ac{KPI} should be established for all buildings in charge. These standards ensure similar operational behavior across different systems. However, flexible operations introduce variable runtime partners. These include fluctuations in operational temperatures, charging/discharging cycles, and intensities~\cite{kumar2020adversarial}. The variations complicate standardized assessment approach and make system diagnosis and management more challenging.\\


Despite the promising outlook of advanced control, significant performance gaps remain in practice between simulation-based environments and real-world outcomes. These gaps stem from epistemic uncertainty due to limited knowledge of systems and aleatory uncertainty due to randomness, such as human behavior. For example, gaps often exist between the anticipated and actual building operations. The high complexity and uncertainty associated with these gaps discourage owners and architects from accounting for operational flexibility in the building design. 
Similar to challenges in the interpretability and transferability of advanced control technologies, performance gaps create discrepancies between perceived and actual benefits. This, in turn, undermines user trust and acceptance~\cite{White2018inaccurate}. \\

New technologies are typically installed incrementally. Once in operation, most building technologies have a lifespan of 10-15 years. When a component reaches the end of its life cycle or malfunctions, only the defective part is replaced, while the rest of the system remains unchanged. This replacement process often focuses only on the technology, overlooking critical aspects such as connectivity and integrated operational capabilities. As a result, high diffusion rates of energy technologies do not necessarily imply greater flexibility in their control. Retrofitting interconnectivity and interoperability is extremely costly~\cite{technology_retrofit}.\\

There is a lack of standardized tests to evaluate the performance and scalability of control strategies across different demand-side resources and applications. Although empirical comparison exists in \cite{natale2023lessons}, a testbed should be designed to accommodate various demand-side resources, control algorithms, market mechanisms, and simulation tools, including representative operational conditions and system configurations. Besides, the control strategies should be capable of optimizing the participation in multiple services, including grid balancing, congestion management, and other grid support services. The control frameworks should be interpretable and transferable, such that they can be adapted to changing market conditions and user preferences to meet regulatory compliance and large-scale rollout.


\subsection*{Economic aspects}
Building a regulatory framework is a time-consuming process and often evolves over the first several years or even decades. As a result, investing in \ac{LFM} solutions brings about external risks for potential stakeholders. These risks are particularly significant when the return on investment expectation spans 8-12 years, which creates barriers to setting up a functional market environment~\cite{sorrell2000reducing}.\\


\ac{LFM} structures vary across regions and countries due to the differences in grid topology, generation sources, customer categorization, consumer need, and policy framework. Service providers and system integrators must be aware of and able to address such specific market settings, which pose a challenge in scaling products and services~\cite{regionality} and maintaining fairness among all market players~\cite{SOTO2021116268}.\\

The existing regulations for trading demand-side flexibility remain insufficient due to the absence of consensus on the market mechanisms for flexibility services. This again raises concerns about social trust, privacy, and fairness. While flexibility markets show promise in managing local congestion or supporting transmission grids, only a few pilot projects currently exist, making it challenging to foresee their development in the coming years~\cite{valarezo2021}. As a result, mechanisms for benefit allocation among stakeholders remain under-investigated. 


\subsection*{Conclusions}
Demand-side flexibility, which leverages electricity consumers' capability to support grid energy supply needs, has emerged as a potent tool for sustainable energy systems in diverse applications.\\

We provide an overview of demand-side flexibility quantification and provision in the building and electric mobility sectors within a socio-techno-economic system. This review underscores the inherent complexity of the problem, necessitating integrating multiple systems and stakeholders spanning societal, technical, and economic dimensions. However, the interdependencies between these dimensions are often overlooked in current studies, limiting the full potential of demand-side flexibility in resources such as buildings and \acp{EV}.\\ 

Addressing these challenges requires interdisciplinary collaboration and coordinated efforts from researchers, policymakers, and industry stakeholders. By enhancing interpretability and transferability, standardizing flexibility KPIs and market mechanisms, and developing comprehensive control mechanisms evaluated against standardized benchmarks, we maximize the potential of demand-side flexibility and benefit efficient, resilient, and sustainable energy systems. Ultimately, demand-side flexibility has the potential to advance the sustainability of future power and energy systems.

\subsection*{Acknowledgements}
We acknowledge contributions from Prof. Jacopo Vivian from Università degli Studi di Padova for the early discussions. H.C. and M.B. disclose support for the research of this work from the nanoverbund project funded by the Swiss Federal Office of Energy (SFOE) (grant number SI/502655). F.B., Y.G., and G.H. disclose support for the research of this work from the SWEET PATHFNDR project funded by the SFOE (grant number SI/502259-01). J.R. and V.N.B. disclose support for the research of this work from NCCR Automation, a National Centre of Competence in Research, funded by the Swiss National Science Foundation (grant number 51NF40\_225155). P.H. and A.C. disclose support for the research of this work from the SWEET Lantern project funded by the SFOE (grant number SI/502544). A.G. discloses support for the research of this work from the European Union’s Horizon Europe (grant number 101138491) and the Swiss Secretariat for Education, Research, and Innovation (grant number 23.00606). L.V.K. discloses support for the research of this work from Empa Forschung \& Entwicklung (grant number 5213.00276). M.M. discloses support for the research of this work from the Swiss Data Science Center (grant number C20-13).

\newpage
\bibliography{bibliography}

\end{document}